\documentclass[prb,twocolumn,superscriptaddress,preprintnumbers,amsmath,amssymb]{revtex4}
\usepackage[dvips]{graphicx}
\usepackage{dcolumn}
\usepackage{color}
\usepackage{bm}

\newcommand{\hatA}{\hat{A}}
\newcommand{\hatrho}{\hat{\rho}}
\newcommand{\hatH}{\hat{H}}
\newcommand{\hatL}{\hat{\mathcal{L}}}
\newcommand{\hatn}{\hat{n}}

\newcommand{\dt}{ \frac{d}{d t}}

\newcommand{\TKKcomp}{Department of Applied Physics/COMP, Aalto University, P.O.~Box 14100, FI-00076 AALTO, Finland}
\newcommand{\TKKltt}{Low Temperature Laboratory, Aalto University, 
  P.O.~Box 13500, FI-00076 AALTO, Finland}

\begin{document}

\title{Cooper pair current in the presence of flux noise}

\author{P. Solinas}
\affiliation{\TKKcomp} 
\affiliation{\TKKltt}

\author{M. M\"ott\"onen}
\affiliation{\TKKcomp}
\affiliation{\TKKltt}
  
\author{J. Salmilehto}
\affiliation{\TKKcomp}

\author{J. P. Pekola}
\affiliation{\TKKltt}

\begin{abstract}
We study the effect of the flux noise on the Cooper pair current and pumping.
We generalize the definition of the current in order to take into account the contribution induced by the environment.
It turns out that this dissipative current vanishes for charge noise but it is finite in general for noise operators that do not commute with the charge operator.
We discuss in a generic framework the effect of flux noise and present a way to engineer it by coupling the system to an additional external circuit. 
We calculate numerically  the pumped charge through the device by solving the master equation for the reduced density matrix of the system and show how it depends on the coupling to the artificial environment.
\end{abstract}

\maketitle 

\section{ Introduction}

Geometric phases are ubiquitous in quantum physics.
Both the Abelian Berry phase \cite{berry84} and the non-Abelian \cite{wilczek84} generalization were theoretically discovered in the eighties.
During the last decade, there has been a renewed interest for their possible use as tools to manipulate quantum information.
The so-called geometric \cite{falci00} or holonomic quantum computation \cite{zanardi99} relies on the fact that it is possible to build unitary transformations which depend only on geometric properties of an abstract parameter space.
The main advantage of this approach is that such geometric transformations are robust against certain types of noise. \cite{dechiara03,solinas04}

While the Berry phase has been observed in many different quantum systems, \cite{abelian_berry,leek07, mottonen08} clear experimental evidence of the non-Abelian adiabatic geometric phases is still missing despite different  proposals for their theoretical implementation. \cite{unanyan99, duan01,fuentes02,recati02,faoro03,solinas03,zangh05}
One of the main reasons for this is that the system must be cyclically steered in an adiabatic fashion. In other words, the time to implement a geometric non-Abelian transformation is long with respect to the dynamical time scale of the system.
This renders it difficult to protect the system from environmental noise and thus decoherence can become an issue in the implementation of geometric quantum computation.

A step towards understanding the effect of the environment and the robustness of steered quantum evolution has been taken in Refs. [\onlinecite{pekola09,solinasPRB10,salmilehto11}] where it has been shown that ground-state evolution is robust against relaxation and dephasing induced by a low temperature dissipative element.

Among different physical systems, the Cooper pair sluice \cite{niskanen03} is a promising candidate for studying the environmental effects on geometric phases.
In fact, the pumped charge through the Cooper pair sluice is known to be directly connected to the geometric phases. \cite{aunola03,mottonen06, brosco08, pirkkalainen10} This has allowed the measurement of the Berry phase  in such a system. \cite{mottonen08}
This has paved the way for ground-state geometric quantum computating in which the quantum information is manipulated through a geometric operator but the system is kept in a doubly degenerate ground state during the whole evolution.\cite{pirkkalainen10, solinasPRA10}

Furthermore, transmon quibits coupled to a superconducting cavity offer potential systems for the observation of the non-Abelian phases. \cite{kamleitnerPRB11}
They constitute the other superconducting platform in which the Berry phase has been observed experimentally.\cite{leek07, neeley09,pechal11}

The Cooper pair sluice operates in the charging regime, i.e., the charging energy dominates over the Josephson energy. For this reason, the noise induced by charge fluctuations is likely to be dominant and it has been already studied in Refs \onlinecite{pekola09} and \onlinecite{solinasPRB10}.
In this paper, we analyze the effect of another kind of noise: the flux noise.
This can be induced, for example, by a fluctuating magnetic field.
We find that the definition of the current must be extended to include the contribution induced by the environment. This dissipative current is directly related to the symmetry of the system and can be calculated using the master equation approach.

As a practical application, we discuss the effect of flux noise on Cooper pair pumping.
Instead of a purely theoretical approach, we study a possible implementation in a realistic experimental setup.
We couple the Cooper pair sluice to an artificial noise source produced by an external circuit.
The main advantage of this approach is the possibility to control in situ the coupling strength between the system and the noise source.

The article is organized as follows.
In Sec. \ref{sec:current_op}, we derive the general expression for the current operator, starting with the emphasis on the symmetries of the problem and then on the connection with the master equation.
In Sec. \ref{sec:pumping},  we apply these results to the pumping process, and identify the different contributions to the current and the pumped charge due to the environment. We analyze the particular case of flux noise which perturbs the phase across the superconducting loop.
Section \ref{sec:eng_env}, presents the circuits to engineer the artificial flux noise environment.
In Sec.\ref{sec:numerics}, we discuss how the pumped charge is influenced by the artificial noise.

\section{Definition of the current operator }
\label{sec:current_op}

Let us assume that a quantum system is in a state described by the density matrix $\hatrho$.
The expectation value of a quantum observable $\hatA$ is
\begin{equation}
 \langle \hatA \rangle={\rm Tr} (\hatrho \hatA) = \sum_{i} \langle i | \hatrho \hatA |i\rangle,
\end{equation}
where the last term is the explicit expression in a time-independent basis $\{ |i\rangle \} $.
If $\hatA$ is time-independent, the time derivative of the expectation value is 
\begin{equation}
\dt\langle \hatA \rangle = \sum_{i} \langle i | \frac{d \hatrho}{d t} \hatA |i\rangle = {\rm Tr} \left( \frac{d \hatrho}{d t}  \hatA \right).
 \label{eq:dotA_initial}
\end{equation}
Using the von Neumann equation we obtain the well-known Ehrenfest theorem
\begin{equation}
 \dt\langle\hatA\rangle = - \frac{i}{\hbar}   {\rm Tr} (\hatrho [\hatA,\hat{H}]),
 \label{eq:dotA}
\end{equation}
where $\hatH$ is the Hamiltonian which determines the dynamics.
Having in mind the usual relation between the charge and the current, we can say that the above equation defines the current associated with the $\hatA$ operator.

The obtained result is valid for a general quantum system.
Let us consider the case in which the total system is composed by a subsystem $S$ and the environment $E$.
We are interested in the dynamics and observables related to the subsystem $S$.
To simplify the discussion, in the following, we will refer to $S$ simply as the system.
The total Hamiltonian can be written as $\hatH= \hatH_S+\hatH_E+\hatH_I$, where
$\hatH_S$ denotes the Hamiltonian of the system, $\hatH_E$ is the Hamiltonian of the environment and $\hatH_I$ describes the interaction between the system and the environment.
The $\hatA$ operator acts only on the system space, i.e., $[\hatH_E,\hatA]=0$, and, from Eq. (\ref{eq:dotA}) we obtain
\begin{equation}
 \dt \langle\hatA\rangle= -\frac{i}{\hbar} \{ \mbox{Tr}(\hatrho [\hatA, \hatH_S ] )+ \mbox{Tr}(\hatrho [\hatA, \hatH_I] ) \}.
 \label{eq:dissipation_definition_first}
\end{equation}
The trace in Eq. (\ref{eq:dissipation_definition_first}) is over all the degrees of freedom and can be split into the trace over the degrees of freedom of the system and the environment: ${\rm Tr}(\hatA)= {\rm Tr}_S[ {\rm Tr}_E(\hatA)]$.
By noticing that both $\hatA$ and $\hatH_S$ act only on the system degrees of freedom and that  ${\rm Tr}_E(\hatrho) = \hatrho_S$ (where $\hatrho_S$ is the reduced density operator of the system), we obtain
\begin{equation}
\dt \langle\hatA\rangle= -\frac{i}{\hbar} \{  {\rm Tr}_S(\hatrho_S [\hatA, \hatH_S ]) + \mbox{Tr}(\hatrho [\hatA, \hatH_I] ) \}.
 \label{eq:dissipation_definition}
\end{equation}

The two contributions in the current associated with $\hatA$ have different origins. 
The first term on the right of Eq. (\ref{eq:dissipation_definition}) resembles the current for an isolated system. 
In fact, if the system does not interact with the environment, $\hatH_I=0$, this is the only contribution present.
The second term can be interpreted as an additional contribution to the current due to the interaction with the environment. 
In the following, we refer to the two contributions as the system current and the dissipative current, respectively.

The case in which $\hatH_I$ commutes with the $\hatA$ operator is particularly interesting.
Here the dissipative contribution to the current vanishes.
However, the interaction with the environment influences the evolution of $\hatrho_S$ and thus it can modify the current. 

\subsection{Connection to the master equation}

Equation (\ref{eq:dissipation_definition}) is written in order to emphasize the symmetry of the problem.
However, if the system and the environment have no particular symmetry, it cannot be used in general to estimate analytically the contribution of the dissipative current since the calculation involves the full density operator.

The dynamics of the reduced density matrix of the system $\hatrho_S$ is obtained by writing the formal solution of the von Neumann equation, expanding it in powers of the system--environment coupling and then taking the trace with respect to the environmental degrees of freedom. \cite{blum} With this procedure we arrive at a standard form of the master equation
\begin{equation}
 \frac{d \hatrho_S}{d t} = -\frac{i}{\hbar} [\hatH_S,\hatrho_S] + \hatL,
 \label{eq:ME}
\end{equation}
where the dissipative term is denoted by $\hatL$.

We can apply this formalism to the case discussed above.
As assumed in the derivation of the master equation, $\mbox{Tr}_E(\frac{d \hatrho}{d t})=\frac{d \hatrho_S}{d t}$ and, thus, $\mbox{Tr}(\frac{d \hatrho}{d t} \hatA) = \mbox{Tr}_S\{ \mbox{Tr}_E (\frac{d \hatrho}{d t} \hatA) \} = \mbox{Tr}_S (\frac{d \hatrho_S}{d t} \hatA)$.
Using this result, Eq. (\ref{eq:dotA_initial}), and Eq. (\ref{eq:ME}), we have
\begin{equation}
  \dt\langle\hatA\rangle	= -\frac{i}{\hbar}  \mbox{Tr}_S( [\hatH_S,\rho_S] \hatA) +   \mbox{Tr}_S(\hatL \hatA). 
 \label{eq:A_ME}
\end{equation}
The first contribution on the right matches with the system current in Eq. (\ref{eq:dissipation_definition}) and, hence, we have an explicit expression for the dissipative current
\begin{equation}
  -\frac{i}{\hbar} \mbox{Tr}(\hatrho [\hatA, \hatH_I] ) = \mbox{Tr}_S(\hatL \hatA).
  \label{eq:dissipative_current_mapping}
\end{equation}
This equation allows us to calculate the dissipative contribution to the current using the expression of $\hatL$
from the master equation for the reduced system dynamics.

Equation (\ref{eq:dissipative_current_mapping}) has important implications involving the quantities that are preserved  during the evolution in presence of environmental noise as recently discussed in Ref. \onlinecite{salmilehto11_short}.

\section{Application to charge pumping}
\label{sec:pumping}

\begin{figure*}
    \begin{center}
    \includegraphics[scale=.7]{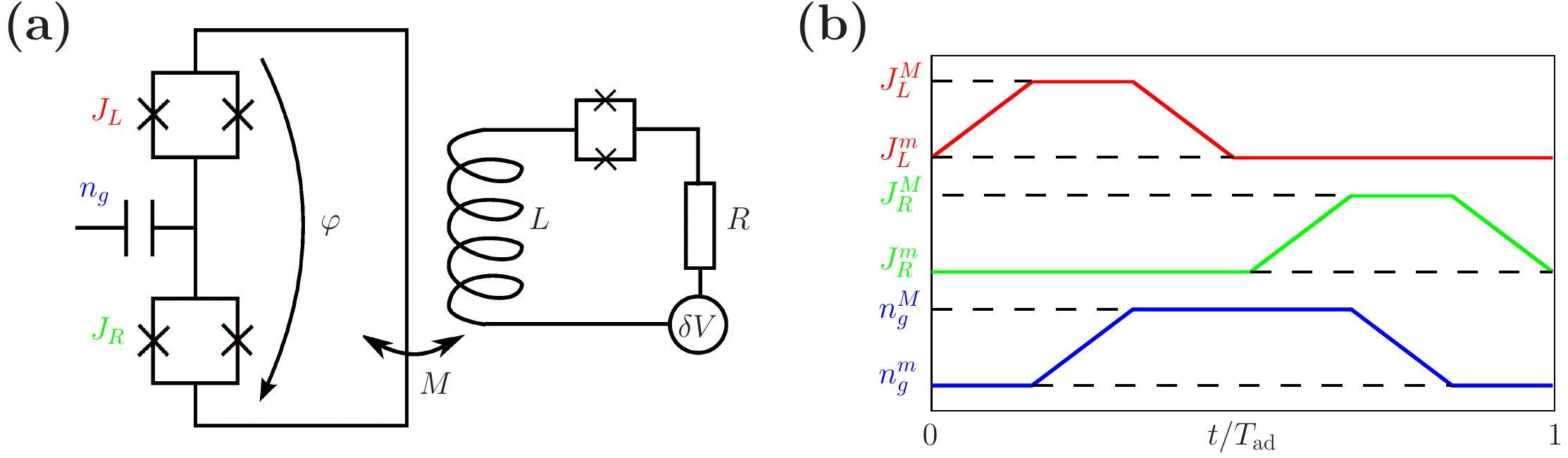}
   \end{center}
    \caption{(Color online) (a) Circuit diagram of the Cooper pair sluice (on the left) and the artificial environment circuit (on the right). The experimentally controlled parameters are $J_{L,R}$ and $n_g$, and $\varphi=2 \pi \Phi/Phi_0$  is the phase difference across the device and $\Phi_0$ is the flux quantum. The system is coupled to the artificial environment through mutual inductance $M$. The engineered environment is  composed of a resistor $R$ associated to a voltage source $\delta V$, an inductor $L$ and an experimentally controllable SQUID. The SQUID allows us to change the effective noise spectrum of the environment.    
    (b) Time dependence of the parameters $J_{L,R}$ and $n_g$ during a Cooper pair pumping cycle.
    }
    \label{fig:system}
\end{figure*}

The above analysis helps us to set the framework to discuss the pumping process in the presence of an environment.
We consider a Cooper pair sluice \cite{niskanen03,niskanen05, vartiainen07} shown in Fig. \ref{fig:system}. It consists of a superconducting island separated by two superconducting quantum interference devices (SQUIDs) with controllable effective Josephson energies $J_{L,R}$.
The electrostatic potential on the island can be controlled by varying a gate voltage $V_g$. 
The experimental access to the parameters $J_{L,R}$ and $V_g$ allows for a full control of the quantum system and makes it an excellent prototype for different applications.
Several steps have been taken with the study of the connection between Cooper-pair pumping and geometric phases, both in its Abelian \cite{mottonen06,mottonen08} and non-Abelian version \cite{brosco08,pirkkalainen10,solinasPRA10}, the robustness of the ground state pumping \cite{pekola09, solinasPRB10, salmilehto10,russomanno11, kamleitner11,salmilehto11} and geometric Landau--Zener--St\"uckelberg interferometry. \cite{gasparinetti11}

The Hamiltonian of the sluice $\hat{H}_S$ is the sum of the charging Hamiltonian $\hat{H}_{\rm ch} =E_C(\hatn-n_g)^2 $ and the Josephson Hamiltonian \cite{niskanen03,niskanen05,pekola09}
\begin{equation}
\hat{H}_J = -J_L \cos \left( \frac{\hat{\varphi}}{2}- \hat{\theta} \right)-J_R \cos \left(\frac{\hat{\varphi}}{2} +\hat{\theta} \right)
 \label{eq:H_J}
\end{equation}
where $\hat{\varphi}=\hat{\varphi}_R+\hat{\varphi}_L$ is the superconducting phase difference between the two leads, 
$n_g=C_gV_g/(2e)$ is the normalized gate charge,  $E_{C}= 2e^2/C_{\Sigma}$ is the charging energy of the sluice, $C_g$ is the gate capacitance, and $C_{\Sigma}$ the total capacitance of the island. 
We denote with $\hatn_k=-i \partial_{\varphi_k} $ ($k=L, R$) the Cooper pair number operator of the $k$th SQUID, and with
$\hat{\theta}=(\hat{\varphi}_R-\hat{\varphi}_L)/2$ and $\hatn= -i \partial_\theta$ the operators for the superconducting phase and the number operator of excess Cooper pairs on the island.
If the device operates in the charge regime, i.e., $E_C \gg \mbox{max}\{J_L,J_R\}=J_{L,R}^M$ and the gate parameter is close to half integer, only the two lowest energy charge states are important for the dynamics and we can adopt the two-state approximation.
Let $|0\rangle$ and $|1\rangle$ denote the states with no and one excess Cooper pair on the island, respectively.
For the system in consideration, it is convenient to reduce the Hilbert space restricting it to states with well-defined $\varphi$.
In this case, $\varphi= 2 \pi \Phi/\Phi_0$ can be treated as a real number and it is determined by the magnetic flux through the large superconducting loop $\Phi$ in Fig. \ref{fig:system}(a).

Starting from the definition of the charge operator through the $k$th SQUID $\hat{Q}_k = - 2 e \hat{n}_k$, the discussion in Sec. \ref{sec:current_op} can be interpreted in terms of physical quantities.
If the dynamics is influenced by the environment, the average current through the $k$th SQUID is
\begin{eqnarray}
 \langle\hat{I}_k\rangle  &=& \frac{d}{dt}\langle\hat{Q}_k\rangle=\frac{2 i e}{\hbar}{\rm Tr}_S (\hatrho_S [\hat{n}_k, \hatH_S]) \nonumber \\ 
 &&+ \frac{2 i e}{\hbar}{\rm Tr}(\hatrho [\hat{n}_k, \hatH_I]). 
 \label{eq:full_current1}
\end{eqnarray}
The current operator for a closed system is usually defined as $\hat{I}_k = \frac{2 i e}{\hbar} [\hat{n}_k, \hatH_S]$ and it corresponds to the first term on the right side in Eq. (\ref{eq:full_current1}) (see Ref. [\onlinecite{mottonen06}]).
Using Eq. (\ref{eq:dissipative_current_mapping}), we have 
\begin{equation}
 \langle\hat{I}_k\rangle  = {\rm Tr}_S (\hatrho_S \hat{I}_k) + \mbox{Tr}_S(\hatL ~\hat{Q}_k).
 \label{eq:full_current2}
\end{equation}
The second contributions in Eqs. (\ref{eq:full_current1}) and (\ref{eq:full_current2}) represent an additional dissipative current 
\begin{equation}
\langle\hat{I}_{k}^{\rm diss}\rangle  = \frac{2 i e}{\hbar}{\rm Tr}(\hatrho [\hat{n}_k, \hatH_I]) = \mbox{Tr}_S(\hatL ~\hat{Q}_k)
 \label{eq:I_diss}
\end{equation}
induced by the environment.

It is convenient to write Eq.~(\ref{eq:full_current2}) in the eigenbasis of $\hatH_S$.
Let $|g\rangle$ and $|e\rangle$ be the eigenstates of $\hatH_S$ and $M_{mn}=\langle m | \hat{M} |n\rangle$ with $m,n=g,e$ for any operator $\hat{M}$. Equation (\ref{eq:full_current2}) becomes
\begin{eqnarray}
 \langle\hat{I}_k\rangle  &=& \rho_{S,gg} I_{k,gg} + \rho_{S,ee} I_{k,ee} +  (Q_{k,gg}-Q_{k,ee}) \mathcal{L}_{gg} \nonumber \\ 
  &+& 2 \Re e (\rho_{S,ge}  I_{k,eg})+ 2 \Re e (\mathcal{L}_{ge} Q_{k,eg}),
 \label{eq:full_current_elements}
\end{eqnarray}
where we have used the fact that, from the symmetries of the master equation, $\mathcal{L}_{ee}=-\mathcal{L}_{gg}$ and $\mathcal{L}_{eg}=\mathcal{L}_{ge}^*$.
We identify the first line as the dynamic current and the second one as the geometric current.
Both are composed of standard contributions $I_{k}^D=\rho_{S,gg} I_{k,gg} + \rho_{S,ee} I_{k,ee}$ and  $I_{k}^G=2 \Re e (\rho_{S,ge}  I_{k,eg})$, and dissipative contributions $I_{k}^{D,\rm diss}=(Q_{k,gg}-Q_{k,ee}) \mathcal{L}_{gg}$ and  $I_{k}^{G,\rm diss}=2 \Re e (\mathcal{L}_{ge} Q_{k,eg})$.
The pumped charge through the $k$th SQUID is defined as
\begin{equation}
 Q_{k}^G  = 2 \int_0^{T_{\rm ad}} \Re e (\rho_{S,ge}  I_{k,eg}) dt +  2 \int_0^{T_{\rm ad}} (\mathcal{L}_{ge} Q_{k,eg}) dt,
 \label{eq:Q_geometric}
\end{equation}
and it is also composed of a standard and a dissipative part.
From the above expressions we can write the current operator for the average current across the device $\hat{I} = (\hat{I}_L +\hat{I}_R)/2$ and the corresponding average geometric current in the sluice.

Note that the dissipative contributions $I_{k}^{D,\rm diss}$ and  $I_{k}^{G,\rm diss}$ are typically small with respect to $I_{k}^D$ and  $I_{k}^G$.
If $\lambda$ is the effective coupling between the system and the environment, the dissipative contribution in the master equation $\hatL$ scales as $\lambda^2$.
Since the master equation (\ref{eq:ME}) is derived in the limit of weak coupling between the system and the environment, we expect a small contribution in Eq. (\ref{eq:full_current_elements}) from the dissipative currents.
However, there can be cases in which,  the dissipative contributions could be detectable if we reduce $I_{k}^{D}$ and  $I_{k}^{G}$.

\subsection{Charge noise environment}
\label{sec:charge_noise}

When the system is in the charge regime, the main source of noise originates from the fluctuations of the gate voltage.
References  [\onlinecite{pekola09, solinasPRB10,salmilehto11}] focus on the study of this charge noise induced by the environment.
If we write the Cooper pair number operator and the charge on the island as $\hatn= \hatn_L - \hatn_R$ and $\hat{Q}= -2 e \hatn$, respectively, the corresponding system--environment interaction is $\hatH_I =-2 e \lambda \hatn \otimes \delta \hat{V} $, where $\lambda$ is the system--environment coupling constant and $\delta \hat{V} $ acts on the environment degrees of freedom. \cite{pekola09, solinasPRB10}
The operators $\hat{n}_k$ commute with $\hatn$ and, thus, there is no dissipative current $\langle\hat{I}_{k}^{\rm diss}\rangle$ through the $k$th SQUID and no total dissipative current contribution.
As discussed in Sec \ref{sec:current_op}, even if the dissipative current vanishes, the total current is modified by the interaction with the environment through the dynamics of $\hatrho_S$ [see Eq. (\ref{eq:full_current1})].


Since no dissipative current can be induced by charge noise, i.e., $[\hat{Q}, \hatH_I]=0$, Eq. (\ref{eq:dissipative_current_mapping}) requires that the term $ \mbox{Tr}_S(\hatL ~\hat{Q})$ calculated from the master equation vanishes.
This can be verified by a direct calculation using the master equations as discussed in Appendix \ref{app:diss_current}.
However, this result depends critically on the form of the master equation used and the approximations done.
In Refs.~\onlinecite{pekola09, solinasPRB10, salmilehto10,salmilehto11} the master equation was obtained keeping the non-secular terms and this procedure produces the expected result.
In a similar way, it can be verified that if we use the secular approximation \cite{blum} and perform the same calculation we obtain $ {\rm Tr}_S(\hatL_{\rm sec} ~\hat{Q}) \neq 0$.
This result is in contradiction with the one based simply on a symmetry argument stating that since $[\hatH_I, \hat{Q}]=0$, we do not have a dissipative contribution to the current.
This observation is enough to state that the secular approximation is not feasible in the description of Cooper pair pumping as has already been pointed out earlier. \cite{pekola09, solinasPRB10}
This result is much more general and can be discussed in completely abstract terms. \cite{salmilehto11_short}

\begin{figure*}
    \begin{center}
    \includegraphics[scale=.7]{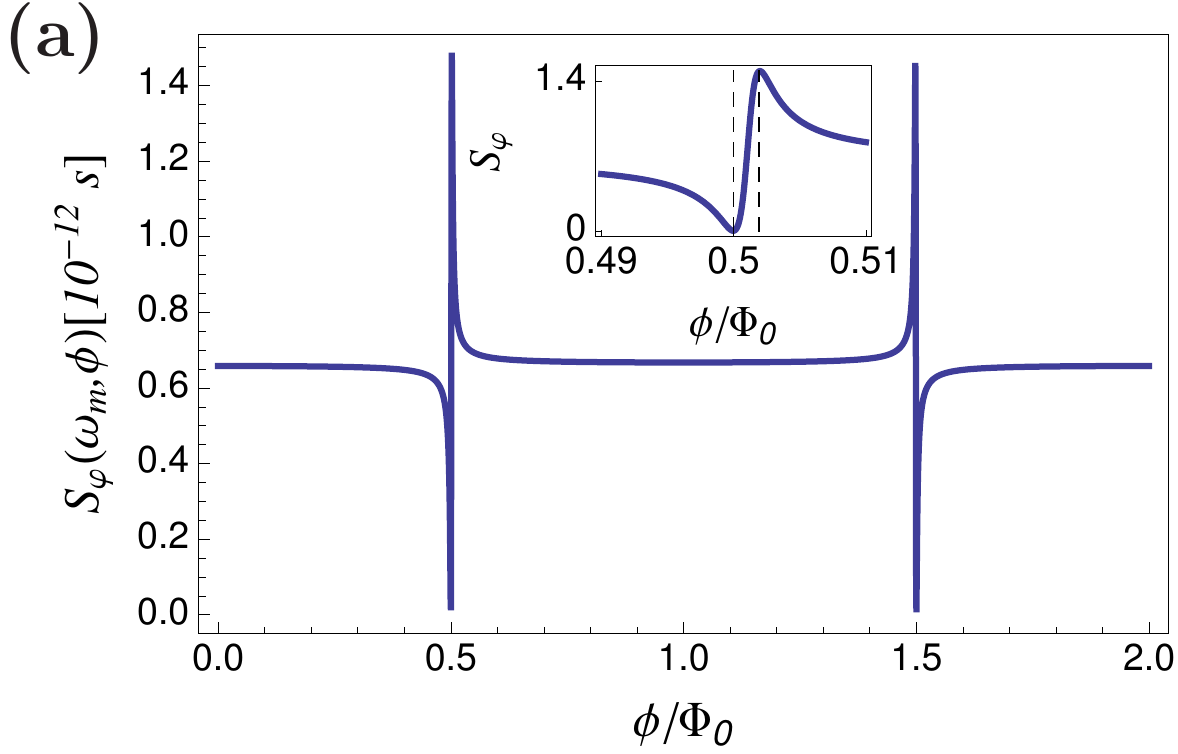}
    \includegraphics[scale=.72]{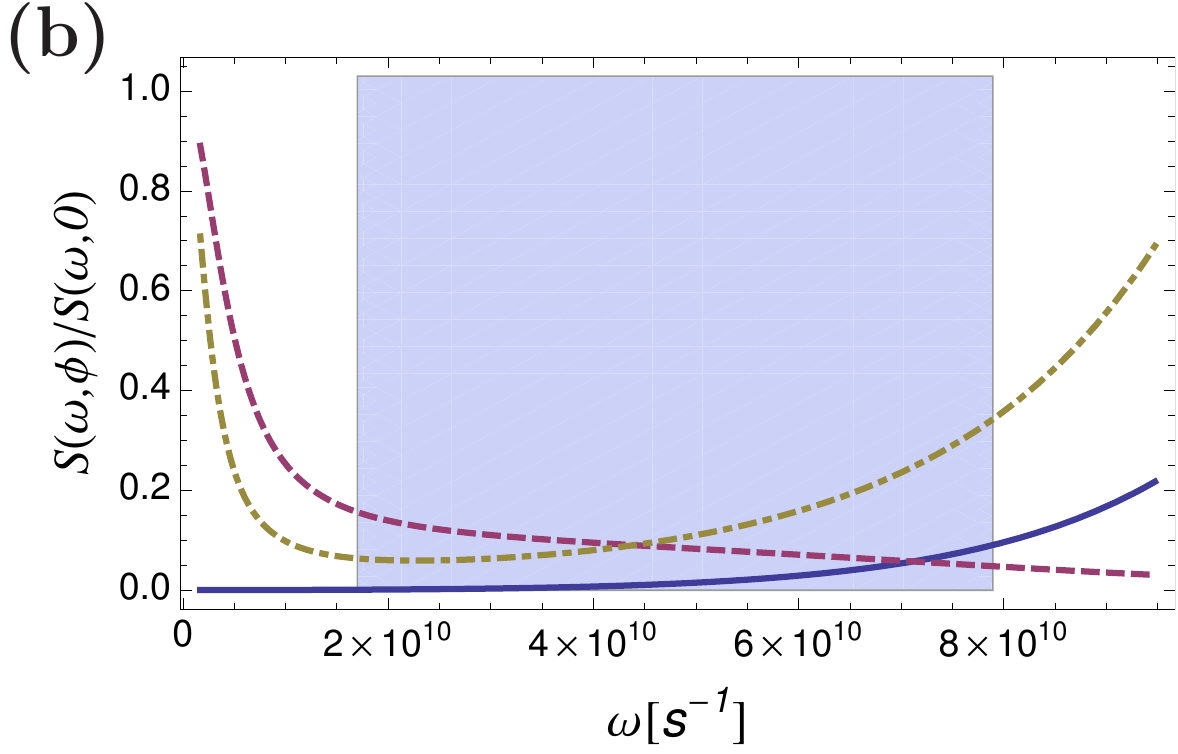}
    \end{center}
    \caption{(Color online) (a) The spectral density function of the phase bias noise as a function of the magnetic flux through the control SQUID $\phi/\Phi_0$ for frequency $\omega_m =1.7 \times 10^{10}$~s$^{-1}$. The inset is a zoom near the resonance point $\phi/\Phi_0=1/2$. The dashed lines show the region delimited by the minimum and maximum $S_\varphi(\omega_m, \phi)$.
    Its width is given by $L_0/(\pi L)$.   
    (b) The ratio $S_\varphi(\omega, \phi)/S_\varphi(\omega, 0)$ for $\phi/\Phi_0=0.5$ (solid line) , $\phi/\Phi_0=0.4991$ (dashed line)  and $\phi/\Phi_0=0.5003$ (dot-dashed line). The shadowed region denotes the frequency range spanned during a typical pumping cycle shown in Fig. \ref{fig:system}(b):  $1.7 \times 10^{10}~$s$^{-1}  \leq \omega \leq  7.9 \times 10^{10}~$s$^{-1}$. 
    The parameters of the artificial noise are $R=30$~$\Omega$, $R_S=500$~$\Omega$, $C_S=50$~fF, $I_C=25$~$\mu$A, $L=M= 0.69$~nH.}
    \label{fig:env_spectrum}
\end{figure*}

\subsection{Flux noise in a Cooper pair sluice}
\label{sec:phase_bias}

If the Cooper pair sluice is subject to flux noise the dissipative contribution to the current is finite and, hence, requires further analysis.
As an example we consider the case in which the fluctuations of the magnetic flux through the outer loop influence the total phase across the sluice $\varphi$. We will refer to this as phase bias noise.

We consider small phase fluctuations in the vicinity of the static point $\varphi_0$ so that the total phase is $\frac{\varphi}{2} = \frac{\varphi_0}{2} +\delta \varphi$ with $ \delta \varphi  \ll \varphi_0$.
The total Hamiltonian is $\hatH_J = \hatH_J(\varphi_0) + \delta \hatH_{J}$ with
\begin{equation}
  \delta \hat{H}_{J}= \left[J_L \sin \left( \frac{\varphi_0}{2}- \hat{\theta} \right)+J_R \sin \left(\frac{\varphi_0}{2} + \hat{\theta} \right) \right] \frac{\delta \varphi}{2}.
\end{equation}

With the two-state approximation, we can write this in terms of the excess of Cooper pairs on the island $\{|0\rangle, |1\rangle \}$. Using the formulas $e^{i \hat{\theta} }= |1\rangle \langle 0|$ and $e^{-i \hat{\theta} }= |0\rangle \langle 1|$, we have
\begin{equation}
 \delta \hat{H}_{J} =  
 \left(\delta J^* |0\rangle \langle 1| + 
 \delta J |1\rangle \langle 0|\right)  \frac{\delta \varphi}{2  },
 \label{eq:dHoperator}
\end{equation}¥
where $\delta J = (\sin \frac{\varphi_0}{2} J_+ +i \cos \frac{\varphi_0}{2} J_-) $ and $J_\pm = J_L \pm J_R$.

Equation (\ref{eq:dHoperator}) can be interpreted as  the system--environment interaction Hamiltonian when $\delta \varphi$ 
 are induced by the environment.
Note that the operator acting on the system degrees of freedom is $\hatH_{I,S}=\delta J^* |0\rangle \langle 1| +  \delta J |1\rangle \langle 0|$. The charge operator on the island is $\hat{Q}= -2e |1\rangle \langle 1|$ and, since $[\hatH_{I,S},\hat{Q}]\neq0 $, we must include the contribution of the dissipative current.
As discussed above, this contribution  can be calculated using the master equation approach. To this end, we can employ the general form of the master equation presented in Ref. [\onlinecite{pekola09}].
The only difference resides in the matrix elements of the coupling operator $\langle m |\hatH_{I,S} | n\rangle $, where $| n\rangle$ and $| m\rangle$ are the time-dependent eigenstates of the sluice Hamiltonian $\hatH_S$, and in the spectral density function of the environment.
The latter is defined as $S_\varphi(\omega) =\int_{-\infty}^{\infty} d \tau \langle \delta \varphi (\tau ) \delta \varphi (0 )\rangle  e^{i \omega \tau}$ and it can be calculated from the correlation function $ \langle \delta \varphi (\tau ) \delta \varphi (0 ) \rangle$ of the phase fluctuations.

\section{Engineered environment for phase bias noise}
\label{sec:eng_env}

To determine experimentally  and discriminate the effect of the environmental noise on the Cooper pair pumping, we should be able to control several properties of the environment. 
This is possible if, in addition to the natural environment, the system is coupled to an engineered source of noise.

A schematic description of the circuit used for the implementation of such a phase bias environment is presented in Fig. \ref{fig:system}.
The main source of noise is a thermal resistor $R$ with noise voltage $ \delta V$.
This circuit is coupled to the system by mutual inductance $M$ and thus it perturbs the phase across the device.
To modify the effects of the environment  we introduce a control SQUID.
By controlling  the flux threading it, we can change the current noise in the circuit which is coupled to the system.

The control SQUID can be represented by a $RLC$ parallel circuit with resistance $R_S$, inductance $L_S$ and capacitance $C_S$.
The impedance of a single SQUID is $Z_{RLC}= \frac{R_S Z_{C_S} Z_{L_S}}{R_S Z_{C_S}+Z_{L_S} Z_{C_S}+R_S Z_{L_S}}= \frac{i L_S(\phi) \omega  R_S}{i L_S(\phi) \omega +R_S \left(1-L_S(\phi) \omega ^2   C_S \right)}$, where $Z_{L_S}$ and $Z_{C_S}$ are the impedances associated to the inductance $L_S$ and capacitance $C_S$, respectively. 
The SQUID inductance $L_S(\phi)= L_0/\cos (\pi  \phi/\Phi_0 )$ can be controlled by changing the normalized flux through the SQUID $\phi/\Phi_0$. The constant component depends on the maximum critical current of the SQUID $I_{C}$:  $L_0=\hbar/(2 \pi e I_{C} )$.
Given  $Z_{RLC}$, we can calculate the total impedance of the circuit $Z_{\rm tot} = Z_{RLC}+  Z_{R}+Z_{L}$ where $Z_R$ and $Z_L$ are the impedances of the resistor and inductor, respectively.

\begin{figure*}
    \begin{center}
    \includegraphics[scale=.75]{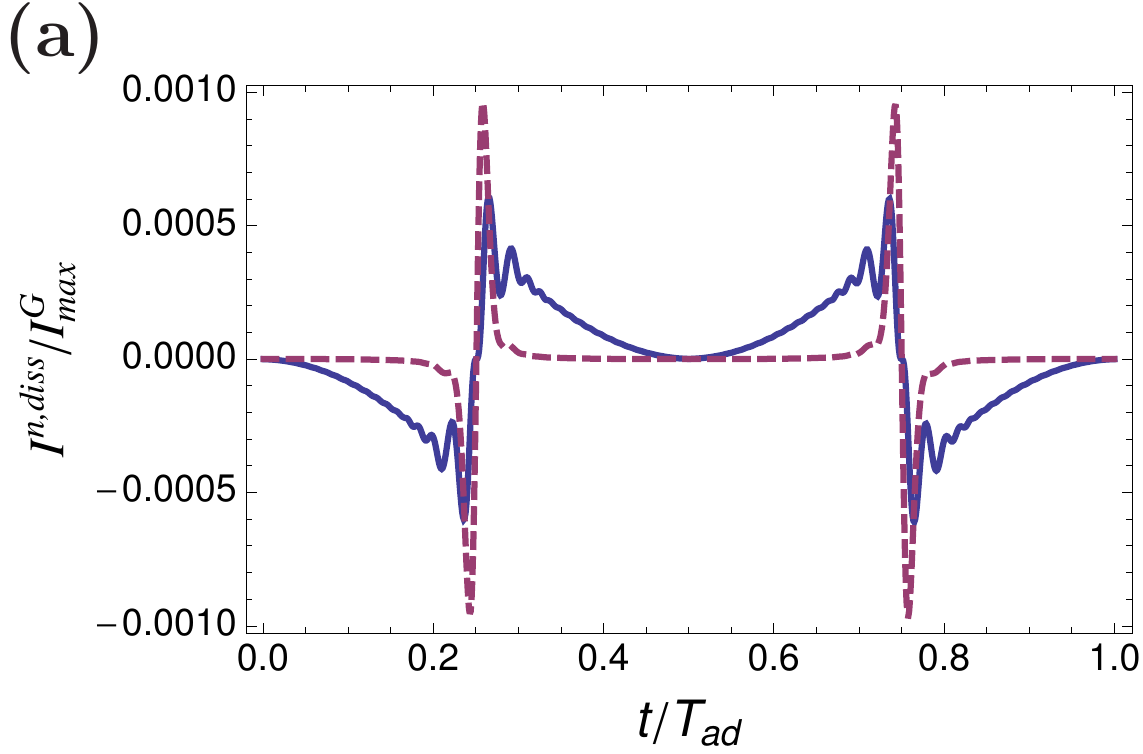}
    \includegraphics[scale=.7]{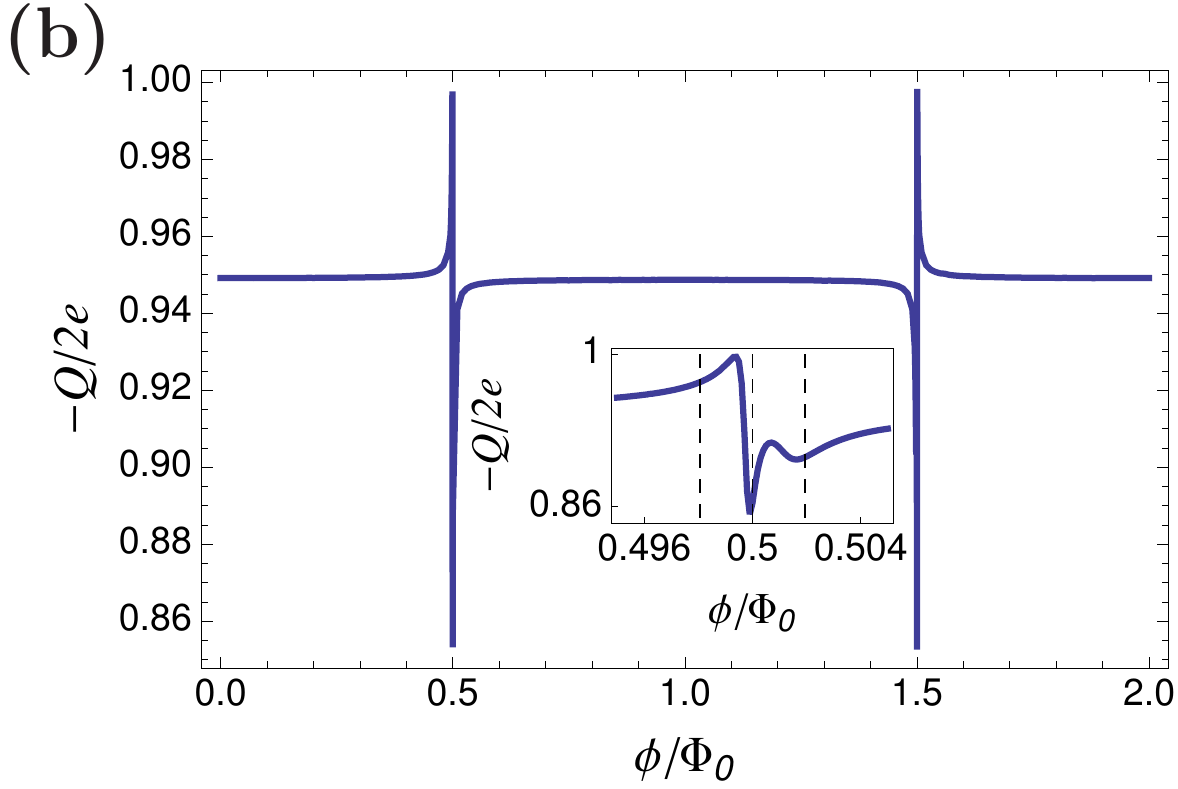}
    \end{center}
    \caption{(Color online)    (a) The dynamic dissipative currents $I^{D,\rm diss}$ (solid line) and the geometric dissipative currents $I^{G,\rm diss}$ (dashed line) normalized to the maximum geometric current $ I^{G}_{max}$ during the pumping cycle. The maximum geometric current during one cycle is $ I^{G}_{max} = 0.5$~nA and the control SQUID of the environmental circuit is open with $\phi/\Phi_0=1$.
    (b) The pumped charge as a function of the flux through the control SQUID of the artificial environment circuit.
    In the inset we show the behavior near one of the resonance points.  
    The dashed lines are determined from Fig. \ref{fig:env_spectrum} (a) and they are located at the minimum $\phi/\Phi_0=1/2$ and at $\phi/\Phi_0=1/2 \pm L_0/(\pi L)$ of the spectral density function.      
    In the numerical simulations, the phase bias is $\varphi_0 = \pi/2$, the pumping frequency is $f=150$~MHz, and the device parameters are $J_i^M/E_{\rm C}=0.1$, $J_i^m/J_i^M=0.03$ (with $i=L,R$), $n_g^M=0.8$ $n_g^m=0.2$, and $E_{\rm C}/k_{\rm B}=1$ K [$E_{\rm C}/(2\pi\hbar)= 21$ GHz].
}
    \label{fig:simulations}
\end{figure*}

We consider a cold resistor with a cut-off frequency much higher than the typical transition energy of the system $\Omega$, i.e., $\Omega \gg k_B T_R$ and $\Omega \ll \omega_c$ where  $\omega_c$, and $T_R$ are the cut-off frequency and the temperature of the noise source, respectively. 
In this limit, the spectral density of the Johnson--Nyquist voltage noise across the resistor is $S_V(\omega)= 2 \hbar \omega R$ 
for $\omega \geq 0$ and $S_V(\omega)=0$ for $\omega < 0$.
We obtain the spectral density of the current noise $S_I(\omega, \phi)=S_V(\omega) / |Z_{\rm tot}|^2$ as 
\begin{widetext}
\begin{equation}
  S_I(\omega, \phi) = \frac{2 R \hbar \omega   }{\left[\frac{\omega ^2 L_S(\phi)^2 R_S}{R_S^2
   \left(1-C_S \omega ^2 L_S(\phi)\right)^2+\omega ^2
   L_S(\phi)^2}+R\right]^2
   +\omega ^2 \left[\frac{L_S(\phi) R_S^2
   \left(1-C_S \omega ^2 L_S(\phi)\right)}{R_S^2 \left(1-C_S
   \omega ^2 L_S(\phi)\right)^2+L_S(\phi)^2 R_S^2}+L\right]^2}.
\label{eq:spectrum_current}
\end{equation}¥
\end{widetext}
The flux noise spectrum is $S_\Phi(\omega, \phi)=M^2 S_I(\omega)$ and, hence, the phase noise spectrum reads
\begin{equation}
  S_\varphi(\omega, \phi) = \frac{ 4 \pi^2 M^2 S_I(\omega, \phi)}{\Phi_0^2}.
  \label{eq:Spectral}
\end{equation}¥
The behavior of $S_\varphi(\omega, \phi)$ as a function of the SQUID flux $\phi$ and for a fixed frequency $\omega_m$ is shown in Fig. \ref{fig:env_spectrum}(a).
The spectrum shows two resonances at $\phi/\Phi_0=1/2,3/2$ and it reflects
the behavior of the control SQUID inductance $L_S(\phi)$.
At these points the artificial environment is maximally decoupled.

It is useful to analyze in detail the behavior of the spectrum near the resonance $\phi/\Phi_0=1/2$.
The spectral density function in Eq. (\ref{eq:Spectral}) for realistic circuit parameters used in the numerical simulations (see Fig. \ref{fig:env_spectrum}) is well approximated by
\begin{equation}
  S_\varphi(\omega_m, \phi) =\frac{8 \pi ^2 M^2 R \hbar \omega_m   }{\Phi _0^2 \left \{ R^2+L^2 \omega_m ^2
   \left[1-\frac{L_0}{\pi L (\phi/\Phi_0-1/2)}\right]^2\right \}}.
   \label{eq:Approx_Spectral}
\end{equation}¥
This expression allows us to calculate the positions and the values of the maximum and the minimum of the spectral density function.
The maximum is found at $\phi_{\rm max}/\Phi_0 = 1/2 + L_0/(\pi L)$ and it is $S_\varphi^{\rm max}(\omega_m, \phi_{\rm max}) = \frac{8 \pi ^2 M^2 \hbar \omega_m   }{R \Phi _0^2}$.
The minimum can be obtained from Eq. (\ref{eq:spectrum_current}) in the limit $\phi/\Phi_0 \rightarrow 1/2$ and it is denoted by $S_\varphi^{\rm min}(\omega_m, 1/2)$.
The width of the drop is given by the difference $\phi_{\rm max}/\Phi_0 -1/2 = L_0/(\pi L)$. It is determined by the ratio between the minimum inductance of the SQUID and the inductance of the circuit [see the inset in Fig. \ref{fig:env_spectrum} (a)].
Using the circuit parameters in Fig. \ref{fig:env_spectrum}, we can estimate the decrease of the noise spectrum at the minimum:
$S_\varphi^{\rm min}(\omega_m, 1/2)/  S_\varphi^{\rm max}(\omega_m, \phi_{\rm max})\} \approx 10^{-4}$ and  $S_\varphi^{\rm min}(\omega_m, 1/2)/  S_\varphi(\omega_m, 0) \approx 10^{-3}$.
This means that we can reduce the strength of the system--environment interaction and effectively decouple the system from the artificial noise source.

The noise influences the circuit at the frequency equal to the energy gap $\Omega$ in the system.
In steered evolution $\Omega$ changes in time and so does the coupling frequency.
To have an estimate of this effect, Fig. \ref{fig:env_spectrum} (b) shows the change in the spectrum as the ratio between $S_\varphi(\omega, \phi)/S_\varphi(\omega, 0)$ for the range of frequencies $\Omega$ spanned in the evolution during the cycle in Fig. \ref{fig:system} (b).
Here controlling the SQUID reduces drastically the strength and the spectral density of the noise during the evolution.

We note that the corresponding decoherence processes induced by the artificial noise are stronger than the ones naturally produced by the flux noise typically observed in the experiment (see Appendix \ref{app:transition_rates}). For this reason, we neglect the effects of the latter and only consider the decoherence induced by the engineered environment.

In the following we also neglect the charge noise.
An estimate of the artificial flux noise strength (see Appendix \ref{app:transition_rates}) shows that, in average, it is stronger than the charge noise except that near the resonance points of $S_\varphi(\omega, \phi)$.
Thus, the effect of artificial flux noise near these regions would be partially hidden.
To observe clearly the influence of the flux noise along for all the values of $\phi$, it is necessary to reduce the charge noise or to further increase the strength of flux noise coupling, i.e., increase the mutual inductance.

\section{Numerical results}
\label{sec:numerics}

Let us study the effect of the artificial noise on the pumped charge in the device shown in Fig. \ref{fig:system}(a). 
The SQUID energies and the gate voltage are modulated within the adiabatic time $T_{\rm ad}=1/f$ where $f$ is the pumping frequency. 
The system is initialized in the ground state and the dynamics is obtained by solving the master equation for the reduced density operator up to the first order in the adiabatic parameter. \cite{pekola09, solinasPRB10} The pumping cycle is repeated until the system reaches the steady state solution $\rho_S(t+T_{\rm ad}) = \rho_S(t)$. We assume that the dissipative source $R$ is at such a low temperature that the environment cannot excite the system.

In Fig. \ref{fig:simulations}(a), we show the time dependence of the dissipative dynamic and geometric currents induced by the phase bias noise.
The curves correspond to the dissipative currents flowing through the circuit when the system reaches the steady state solution. 
The dissipative currents are about three orders of magnitude smaller than the total geometric current.
The corresponding dissipative pumped charge is almost zero.
This is not surprising since we are analyzing the system in the steady state solution, and since the dissipative currents have no favored direction, the corresponding charge averages out over many pumping cycles. 
The dissipative contributions are significantly greater during the first pumping cycles when the system is still far from the steady state.

In Fig. \ref{fig:simulations}(b), we present the stationary pumped charge under the influence of the artificial flux noise. The tuning parameter is the flux of the control SQUID $\phi/\Phi_0$ which allows us to effectively change the system--environment coupling.
As shown, the pumped charge is almost constant except near the resonance points $\phi/\Phi_0=1/2,3/2$.
The global behavior of the pumped charge and its drop at the resonance points can be physically explained keeping in mind that here the system--environment coupling is effectively suppressed [Fig. \ref{fig:env_spectrum}(a)].
The stationary pumped charge is determined by two competing effects.
The non-adiabatic transitions tend to excite the system while the relaxation induced by the environment tends to keep the system in the ground state. Since the ground and excited states pump in opposite directions, the first effect reduces the pumped charge while the second one stabilizes it near to the expected value of one Cooper pair per cycle. \cite{pekola09, solinasPRB10, note_firstorder_MEQ}
The stationary solution is reached after many pumping cycles when the two effects are in balance.
Away from the resonance points, the spectrum is almost constant producing a constant pumped charge.
At the resonances, the results in Fig. \ref{fig:simulations}(b) reflect the fact that, since we are effectively decreasing the system--environment coupling, the non-adiabatic excitations become more important.
Correspondingly, the excited-state component is enhanced  in the final steady state  and we observe a smaller pumped charge.
The maximum drop of the pumped charge  is $(Q_{\rm max}-Q_{\rm min})/Q_{\rm max} \approx 15 \%$ and the drop from the pumped charge away from the resonance  $(Q_0-Q_{\rm min})/Q_0 \approx 10 \%$ [$Q_0= Q(\phi=0)$].

However, the details of the pumped current in the proximity of the resonances are different from the ones exhibited by the environment spectrum $S_\varphi(\omega, \phi)$. This reflects the fact that the pumped charge is an observable that can be influenced by many different effects.
In particular, the pumped charge first increases near the resonance points  before decreasing. 
It has clearly s non-trivial behavior [inset in Fig. \ref{fig:simulations} (b)].
One possible explanation of this behavior can be the influence of the Landau--Zener--St\"uckelberg (LZS) interference. \cite{gasparinetti11} During the pumping cycle in Fig. \ref{fig:system}(b), the system crosses two avoided crossing where $n_g=1/2$.
Near these points Landau--Zener transitions occur and the phase difference between the excited and the ground state accumulated in the intermediate region leads to interference effects. The overall result is to change the ground-state population and thus the pumped charge.
In the present model, the phase bias is kept constant but the artificial noise can effectively induce a change in the accumulated phase difference and hence in the LZS interference.
If the system is initially in the ground state, the probability to excite it due to LZS interferometry after one cycle  is $\mathcal{P}_e = P_{LZ} (1-P_{LZ}) \cos^2 (\alpha + \varphi/2)$ where 
$P_{LZ}$ is the Landau--Zener transition probability and $\alpha$ depends on the energy gap and on the pumping frequency.  \cite{gasparinetti11}
If $\delta \varphi$ is the perturbation on the constant phase bias $\varphi_0=\pi/2$, we have $\varphi/2= \pi/4 + \delta \varphi$.
The average of the excitation probability during a loop is 
\begin{equation}
 \mathcal{P}_e \approx \mathcal{P}_e^0 + \frac{1}{2} P_{LZ} (1-P_{LZ})   \sin (2 \alpha) \langle \delta \varphi^2\rangle,
\end{equation}
where $\mathcal{P}_e^0 $ is the excitation probability without noise, $\langle \delta \varphi^2\rangle$ is the average of the square phase fluctuation and  we have assumed $\langle\delta \varphi\rangle=0$.
We calculated the contribution due to the noise as $\langle \delta \varphi^2\rangle= \int_{\omega_{m}}^{\omega_{M}} S_\varphi(\omega, \phi) d \omega$ and, as a function of the control flux $\phi$, and observe that it behaves as $S_\varphi(\omega, \phi)$ as a function of $\phi$.
The probability to remain in the ground state is $\mathcal{P}_g =1- \mathcal{P}_e$ and its dependence on the control SQUID flux $\phi$ is inverted with respect to $S(\omega, \phi)$.
This means that when the strength of the noise spectrum increases because of the LZS effect, the ground state is less populated and the pumped charge decreases.
The final behavior of the pumped charge as a function of $\phi$ is determined by the noise spectrum, the LZS interference and their relative strength integrated over the pumping cycles.
 
Note that for the numerical simulations we have used the master equation which includes only the first order contribution in the perturbative adiabatic parameter. \cite{pekola09,solinasPRB10}
To obtain a more accurate value of the pumped charge a different master equation which includes high-order corrections can be used. \cite{salmilehto10,salmilehto11}

\section{Conclusions}
\label{sec:conclusions}

In summary, we have presented a study of the flux noise on the Cooper pair pumping process.
Firstly, we showed that in the presence of flux noise the current operator must be modified in order to take into account the current induced by the environment.
This dissipative current is related to the symmetry of the system and can be calculated from the master equation for the reduced density matrix.

Secondly, we then analyzed the effects of the phase bias noise on the Cooper pair pumping.
In the model studied,  the noise is produced by an artificial environment coupled to the system by mutual inductance.
The advantage of this scheme is that, introducing a control SQUID in the environmental circuit, we have access to the system--environment coupling and can essentially decouple the system from the environment.
The pumped charge obtained by solving the master equation shows clearly the features induced by the noise.

The system presented here operates in the charge regime and it is then primarily sensitive to charge noise.
For this reason, it can be challenging to experimentally measure the predicted effect because it can be partially hidden by the charge noise. However, there are several modifications which can increase the experimental accessibility:
reducing the effect of charge noise increasing the mutual inductance coupling with a different design of the artificial noise circuit. For example, using multilayer lithographic techniques, it is possible to increase the coupling between the system and the environment.

This work is a step towards understanding the effect of different types of noises on steered superconducting devices.
Such understanding is critical for practical applications such as metrology and quantum information processing.
A deeper knowledge of the effect of the environment can open a way to the design and implementation of robust devices or, in the case of quantum information, error correction techniques.

In this direction, the pioneering experimental works of implementing  engineered environments have been carried out using trapped ions. \cite{myatt00, kielpinki01} One of the most striking results of these experiments was the proof that it is possible to store  quantum information in states which are robust against decoherence. \cite{kielpinki01}
Still, apart from a few theoretical  proposals, \cite{solinasPRB10} similar experiments are missing in condensed matter systems.

Along these lines, the next step is to analyze the effect of similar flux noise sources on superconducting devices in different regimes. For example, in the transmon regime the system is insensitive to charge noise and the flux noise becomes the dominant source of noise.\cite{koch07}
Thus, using the discussed engineered environment, we should be able to study and measure the environmental effect in controlled situations.

\section*{Acknowledgements}

We have received funding from the European Community's Seventh Framework Programme under Grant Agreement No. 238345 (GEOMDISS). We acknowledge Academy of Finland, Emil Aaltonen Foundation,
V\"ais\"al\"a Foundation, and the KAUTE Foundation for financial support.

\appendix
\section{Absence of dissipative current with charge noise environment}
\label{app:diss_current}

By using Eq. (\ref{eq:dissipative_current_mapping}) we can check by a direct calculation that, when the system is subjected to charge noise, there is no dissipative current contribution.
When the system is adiabatically steered along a cyclic loop, the master equation is conveniently written in the time dependent eigenbasis of $\hatH_S$.
There are several equivalent approaches \cite{pekola09, solinasPRB10, salmilehto10,salmilehto11} exploiting the two-level approximation in the weak coupling limit.
Since the present calculation can be done in an abstract way without explicitly taking into consideration the form of $\hatH_S$, it is convenient to use the master equation in the superadiabatic basis representation \cite{salmilehto10,salmilehto11} which is easier to manipulate.

Let $\hat{D}_1$ be the operator which diagonalizes $\hatH_S$ in a fixed basis.
Then the dynamics of the closed system are effectively governed by $\hat{H}_{S,1} = \hat{D}_1^{\dagger}\hat{H}_S\hat{D}_1 + \hbar \hat{w}_1$ where $\hat{w}_1 = -i\hat{D}_1^{\dagger}\dot{\hat{D}}_1$. 
Defining $\hat{D}_2$ as the diagonalizing operator for $\hat{H}_{S,1}$ allows us to define a new effective Hamiltonian as $\hat{H}_{S,2} = \hat{D}_2^{\dagger}\hat{H}_S\hat{D}_2 + \hbar \hat{w}_2$ where $\hat{w}_2 = -i\hat{D}_2^{\dagger}\dot{\hat{D}}_2$.
If the evolution is sufficiently slow, that is, $|\hat{w}_2| \ll |\hat{w}_1|$ we may approximate $\hat{H}_{S,2} \approx \hat{D}_2^{\dagger}\hat{H}_S\hat{D}_2$ yielding an effectively nonsteered picture. 
Denoting the superadiabatic basis as 
$\{  |g\rangle = \hat{D}_1\hat{D}_2 |0\rangle,  |e\rangle = \hat{D}_1\hat{D}_2|1\rangle \}$, where $\{ |0\rangle, |1\rangle \}$ defines the fixed basis, yields a master equation in the Schr\"odinger picture as
\begin{equation}
\frac{d\rho_{gg}}{dt} = - (\Gamma_{ge}+\Gamma_{eg})\rho_{gg} + \Re e\{ \tilde{\Gamma}_0\rho_{ge} \} + \Gamma_{eg},
\label{eq:rho_gg1}
\end{equation}
and 
\begin{equation}
\begin{split}
\frac{d\rho_{ge}}{dt} &= i\Omega\rho_{ge} - (\tilde{\Gamma}_++\tilde{\Gamma}_-)\rho_{gg} \\ &- \bigg( \frac{\Gamma_{eg}}{2}+\frac{\Gamma_{ge}}{2}+\Gamma_{\varphi} \bigg) \rho_{ge} + (\Gamma_{\alpha}+\Gamma_{\beta})\rho_{eg} + \tilde{\Gamma}_+,
\end{split}
\label{eq:rho_ge}
\end{equation}
where $\rho_{rs} =\langle r|\hat{\rho}_S|s \rangle$ with $r,s \in \{g,e\}$, and $\Omega = (E_e-E_g)/\hbar$. 
If $\hatH_I= \hat{Z}\otimes \hat{X}$ where $\hat{Z}$ and $\hat{X}$ are the operators acting on the system and the environment, respectively, the rates in Eqs. (\ref{eq:rho_gg1}) and (\ref{eq:rho_ge}) are defined as 
\begin{gather}
\Gamma_{ge} = \frac{|\langle e|\hat{Z}|g\rangle|^2}{\hbar^2} S(-\Omega), \nonumber \\ 
\Gamma_{eg} = \frac{|\langle e|\hat{Z}|g\rangle|^2}{\hbar^2} S(+\Omega), \nonumber \\
\tilde{\Gamma}_0 = \frac{\langle e|\hat{Z}|g\rangle( \langle g|\hat{Z}|g\rangle-\langle e|\hat{Z}|e\rangle)}{\hbar^2} S(0), \nonumber \\
\tilde{\Gamma}_{\pm}=\frac{\langle g|\hat{Z}|e\rangle(\langle  e|\hat{Z}|e\rangle-\langle g|\hat{Z}|g\rangle)}{2\hbar^2}S(\pm\Omega),  \\
\Gamma_{\varphi}=\bigg( \frac{|\langle e|\hat{Z}|e\rangle|^2}{2\hbar^2} + \frac{|\langle g|\hat{Z}|g\rangle|^2}{2\hbar^2} - \frac{\langle g|\hat{Z}|g\rangle\langle e|\hat{Z}|e\rangle}{\hbar^2} \bigg) S(0), \nonumber \\
\Gamma_{\alpha}=\frac{\langle g|\hat{Z}|e\rangle^2}{2\hbar^2}S(\Omega), \nonumber \\
\Gamma_{\beta}=\frac{\langle g|\hat{Z}|e\rangle^2}{2\hbar^2}S(-\Omega). \nonumber 
\label{eq:rates}
\end{gather}
The spectral density is denoted by $S(\omega) = \int_{-\infty}^{\infty} d\tau \langle \hat{X}(\tau) \hat{X}(0)\rangle e^{i\omega \tau}$. 
Note that Eqs. (\ref{eq:rho_gg1}) and (\ref{eq:rho_ge}) still contain the non-secular terms.

We can write the superadiabatic states in a general form as
\begin{eqnarray}
  |g\rangle &=& \cos \theta |0\rangle + \sin \theta e^{i \phi} |1\rangle, \nonumber \\
  |e\rangle &=& \sin \theta |0\rangle - \cos \theta e^{-i \phi} |1\rangle, 
  \label{eq:ge_basis}
\end{eqnarray}
where $\theta$ and $\gamma$ are time dependent functions.
We consider a generic diagonal noise operator in the charge basis: $\hat{Z}=a |1\rangle \langle 1|+b|0\rangle \langle 0|$ for $a, b \in \mathbb{R}$.
Expressing the charge operator of the island as $\hat{Q} = -2 e  |1\rangle \langle 1|$ and the noise operator $\hat{Z}$ in the time-dependent basis (\ref{eq:ge_basis}), we have
\begin{eqnarray}
\hat{Q}&=&-2 e [ \sin ^2\theta |g\rangle \langle g| + \cos ^2\theta  |e\rangle \langle e| \nonumber \\ 
 &&-\cos \theta  \sin \theta ( |e\rangle \langle g| +  |g\rangle \langle e|) ],
\end{eqnarray}
and 
\begin{eqnarray}
\hat{Z}&=& (a \cos ^2\theta +b \sin ^2\theta) |g\rangle \langle g|  \nonumber \\ 
           && + (b \cos ^2 \theta +a \sin ^2\theta)  |e\rangle \langle e|  \nonumber \\ 
 &&(a-b) \cos \theta  \sin \theta  ( |e\rangle \langle g| +  |g\rangle \langle e|) .
\end{eqnarray}
At the same time the $\mathcal{L}_{ij}$ terms in Eq. (\ref{eq:A_ME}) can be read directly from the dissipative part of the master equation 
\begin{eqnarray}
  \mathcal{L}_{gg} &=&  - (\Gamma_{ge}+\Gamma_{eg})\rho_{gg} + \Re e\{ \tilde{\Gamma}_0\rho_{ge} \} + \Gamma_{eg}  \nonumber \\
\mathcal{L}_{ge} &=& - (\tilde{\Gamma}_++\tilde{\Gamma}_-)\rho_{gg} - \bigg( \frac{\Gamma_{eg}}{2}+\frac{\Gamma_{ge}}{2}+\Gamma_{\varphi} \bigg) \rho_{ge}  \nonumber \\ && + (\Gamma_{\alpha}+\Gamma_{\beta})\rho_{eg} + \tilde{\Gamma}_+.
\end{eqnarray}
Thus we have all the elements to calculate the dissipative current 
\begin{equation}
 \mbox{Tr}_S (\hatL ~\hat{Q})= (Q_{gg}-Q_{ee}) \mathcal{L}_{gg} +  2 \Re e (\mathcal{L}_{ge} Q_{eg}).
 \label{eq:L_contribution}
\end{equation}
 By explicitly writing the transition rate in Eq. (\ref{eq:rates}) for the noise operator $\hat{Z}$ and inserting all the contributions in Eq. (\ref{eq:L_contribution}), we verify that ${\rm Tr}_S(\hatL ~\hat{Q})=0$.
These results are valid for any perturbation operator diagonal in the $\{  |0\rangle ,  |1\rangle  \}$ basis and thus for the charge noise discussed in Sec. \ref{sec:charge_noise}.
The same result can be obtained using the master equation in the adiabatic basis\cite{pekola09, solinasPRB10}.
 
It is interesting that the condition $ {\rm Tr}_S(\hatL ~\hat{Q})=0$ depends critically on the form of the master equation.
As an example, we consider the same problem but we perform the usual secular approximation.\cite{blum}
The correctness of the secular approximation in the analysis of the charge pumping has been questioned \cite{pekola09, solinasPRB10} since it has been shown that it can lead to unphysical results such as charge non-conservation.
The master equation after the secular approximation reads 
\begin{equation}
\frac{d\rho_{gg}^{\rm sec}}{dt} = - (\Gamma_{ge}+\Gamma_{eg})\rho_{gg}^{\rm sec}  + \Gamma_{eg},
\end{equation}
and 
\begin{equation}
\frac{d\rho_{ge}^{\rm sec}}{dt} = i\Omega\rho_{ge}^{\rm sec}  - \bigg( \frac{\Gamma_{eg}}{2}+\frac{\Gamma_{ge}}{2}+\Gamma_{\varphi} \bigg) \rho_{ge}^{\rm sec}.
\end{equation}
In this case, it can be shown that a calculation similar to the one above gives non-vanishing dissipative current since 
$ {\rm Tr}_S(\hatL_{\rm sec} ~\hat{Q}) \neq 0$.
This result is in contradiction to the one based on the symmetry argument.
Similar results can be obtained in a more formal framework as discussed in Ref. \onlinecite{salmilehto11_short}.

\section{Artificial and natural decoherence rates}
\label{app:transition_rates}
In addition to the artificial environment discussed in Sec. \ref{sec:eng_env}, we should also take into account the effect of the natural flux noise. 
It is sufficient to consider the generic decoherece rate
$\Gamma(\phi) = |\langle i|\delta \hatH | j \rangle/\hbar|^2 S_\varphi(\omega,\phi)$ where $S_\varphi(\omega,\phi)$ is the spectral density function of the environment.
The matrix element $|\langle i| \delta \hatH | j \rangle|^2$ is associated to the relaxation process if $i \neq j$ and the dephasing if $i=j$.
The order of magnitude of $ |\langle i|\delta \hatH | j \rangle/\hbar|^2$ for the pumping cycle in Fig. \ref{fig:system}(b) can be estimated using Eq.~(\ref{eq:dHoperator}) as $(J_M/\hbar)^2$.
Combining this with the values of $S_\varphi(\omega,\phi)$, we have an estimate of the decoherence times (both dephasing and relaxation), $\tau (\phi) = 1/ \Gamma(\phi)$.

During the pumping cycle the system energy gap changes in time \cite{pekola09, solinasPRB10} and the minimum energy gap is reached near the degeneracy points where $n_g=1/2$. For the majority of the evolution time, the system frequency is close to its maximum $\omega_M=7.9 \times 10^{10}$~s$^{-1}$. We can use this reference frequency to estimate the decoherence rates.
The minimum and maximum values of the spectral density function at $\omega_M$ are $S_{\varphi}^{\rm min} = 2.8 \times 10^{-15}$~s and $S_{\varphi}^{\rm max} = 3 \times 10^{-12}$~s.
The estimated relaxation and depahsing times with the parameters used in the numerical simulation (see the caption of Fig. \ref{fig:simulations}) are $\tau_{\rm min} \approx 1.9$~ns and $\tau_{\rm max} \approx 2~\mu$s.

For the charge noise, the decay and dephasing times to the can be estimated as in Ref. \onlinecite{pekola09} and they are of order $10$~ns. Thus, the artificial flux noise is the dominant source of noise except  that near the resonance points of $S_\varphi(\omega, \phi)$.


The effect of the natural environment can be estimated from the experimental results of the decoherence time for devices affected by flux \cite{itier05,yoshihara06,bylander11}.
We focus on the effect of ubiquitous $1/f$ noise.
In this case, the measurements of the decoherence induced by such a noise suggest that the decay has Gaussian and not exponential shape \cite{yoshihara06,bylander11}.
Note that, strictly speaking, this implies that it cannot be described by a linear master equation. However, a phenomenological approach has been applied \cite{itier05}. 
The effective dephasing rate is $\Gamma = \sqrt{A_{\varphi} \ln2} |\partial \Omega/ \partial \varphi|$ where the spectral density function for the $1/f$ noise is $S_{\varphi}(\omega) = A_{\varphi}/|\omega|$  and $\Omega$ is the frequency of the system.
The maximum of $|\partial \Omega/ \partial \varphi|$ is obtained when the system is at the degeneracy point, i.e., $n_g=1/2$ and only the Josephson contribution is present Hamiltonian [see Eq. (\ref{eq:H_J})].
In this case, $\Omega= \sqrt{J_R^2+J_L^2+2 J_R J_L \cos \varphi}/\hbar$ and we can approximate $J_R\approx J_M \gg J_L \approx J_m$.
From the measured amplitude of the magnetic flux spectrum $A_{\Phi} = (1.7~\mu \Phi_0)^2$, we can calculate the amplitude of the phase spectrum $A_{\varphi} =  4 \pi^2 A_{\Phi}/\Phi_0^2$.
Thus, we obtain the dephasing time due to low-frequency noise: $\tau \approx 28~\mu$s.
This dephasing time is very long compared to the artificial decoherence time, and hence the natural low-frequency noise does not change the results of the presented numerical simulations.


\begin{thebibliography}{99}
\bibitem{note_firstorder_MEQ} 
We recall that we are using a master equation at the first order in the adiabatic parameter.
A more accurate prediction for the pumped charge when the system--environment coupling increases can be obtained including the higher order corrections.
\end{thebibliography}
\end{document}